\documentstyle[12pt]{article}

\begin{document}

\title{On the Security of Quantum Cryptography 
          Against Collective Attacks}

\author{Eli Biham$^{(1)}$ and Tal Mor$^{(2)}$}

\date{\today}

\maketitle

{(1)~Computer Science Department, Technion, Haifa
32000, Israel; 
(2)~Physics Department, Technion, Haifa 32000, Israel;}

\begin{abstract}

We present strong attacks against quantum key distribution schemes which
use quantum memories and quantum gates to attack 
{\em directly} the final key.
We analyze a specific attack of this type, for which 
we find the density matrices available to the eavesdropper and the 
optimal information 
which can be extracted from them.
We prove security against this attack 
and discuss security against any attack allowed by the rules of 
quantum mechanics.
\newline
\newline
{PACS number(s): 03.65.Bz, 89.70, 89.80}
\end{abstract}

Quantum cryptography~\cite{BB84,Ben92,EPR,BBBSS,BHM} uses quantum mechanics 
to perform new cryptographic tasks --- especially 
information secure key distributions --- which are beyond the abilities of
classical cryptography.
Unfortunately, the security of such a key  
is still unproven: 
Sophisticated attacks (called {\em coherent} or {\em joint attacks})
which are directed against the final key were suggested; 
The analysis of such attacks is very complicated, 
and, by the time this work was submitted, 
security against them was proven only in the non-realistic case 
of ideal (error-free) channels~\cite{Yao,DM3}.
The security in the real case, which is crucial for making quantum
cryptography practical,  
is commonly believed but yet unproven. 
A proof of security must bound the information available to
the eavesdropper (traditionally called Eve), 
on the final key, to be negligible
(i.e., much smaller than one bit).
A protocol is considered secure if 
the adversary is restricted only by the rules of quantum mechanics,
and a protocol is considered practical if the legitimate users are
restricted to use existing technology. 
In this work we obtain the strongest security result for practical protocols.
We suggest {\em collective attacks} (simpler than the joint attacks)
which are simple enough to 
be analyzed, but are general enough to imply (or at least suggest) the security
against any attack. 
We prove security against the simplest 
collective attack:
we generalize methods developed in~\cite{BMS} in order to calculate Eve's 
density matrices explicitly, and to find 
the information which can be obtained from them; we show that it 
is negligible.
Our result also provides better understanding of the issue of information
splitting between two parties which is a fundamental
problem in quantum information theory.   
Parts of this work were done together with Dominic Mayers.

In any quantum key distribution scheme, the sender, Alice,
sends to the receiver, Bob, a classical string of bits by encoding them
as quantum states.
In the two-state scheme~\cite{Ben92} (B92 scheme)
a classical bit is represented by either of 
two non-orthogonal pure states, which can be written as  
$ \psi_0 = {\cos \theta \choose  \sin \theta} $, and 
$ \psi_1 = {\cos \theta \choose - \sin \theta} $.
Bob performs a test which provides him with a conclusive or inconclusive
result. For instance, he can test whether a specific particle is in a state
$\psi_0$ or a state orthogonal to it ${\psi_0}'$; A result $\psi_0$ is 
treated as inconclusive 
and a result ${\psi_0}'$ is identified as $\psi_1$.
Alice and Bob use also an unjammable classical channel to inform
which bits were identified conclusively, and 
to compare some of the common bits in order to estimate the error-rate.
They must accept some small error-rate $p_e$ due to imperfections in 
creating, transmitting and receiving of the quantum states.
If the estimated 
error-rate exceeds the allowed error-rate
they quit the transmission
and do not use the data, thus any eavesdropping attempt is severely
constrained to induce an error-rate smaller than $p_e$.
Alice and Bob are now left with similar  $n$-bit strings which contain errors. 
They randomize the order of the bits and correct the errors
using any error-correction code~\cite{MS}.
The error-correction code is usually made of $r$ parities of substrings
(where the parity bit $p(x)$ of a binary string $x$ is zero if there is even
number of $1$'s in $x$, and one otherwise).
Alice sends these parities to Bob (using the classical channel), 
who uses them to obtain a (possibly shorter) string 
identical to Alice's, 
up to an exponentially small error probability. 
Finally, Alice and Bob can amplify the security of
the final key by using privacy amplification techniques~\cite{BBCM}: 
by choosing some parity bits
of substrings to be the final key. 
Their aim is to derive a final key on which Eve's average information
is negligible.

Eve can measure some of the particles and gain a lot of information on them,
but this induces a lot of error. Hence, she can attack only  
a small portion of the particles, and this 
reduces her information
on the parity of many bits exponentially to zero. 
Translucent attacks~\cite{EHPP}  
are much more powerful:
Eve attaches a probe to {\em each particle} 
and performs some unitary transformation,
after which her probe is correlated to the transmitted state.
In the case where each probe is left in a pure state~\cite{EHPP},
and measured separately to obtain information on Alice's bit,
it is a rather obvious conclusion (from~\cite{BBCM}) 
that privacy amplification is still effective.
Thus, such an {\em individual} translucent attack is ineffective.
We deal with a much more sophisticated attack
in which Eve's measurement is done {\em after}
the processes of error-correction and privacy amplification are completed.
Privacy amplification techniques were not designed to stand against such 
attacks, hence their efficiency against them is yet unknown.
Consider the following {\em collective} attack:
(1) Eve attaches a \underline{separate, uncorrelated} 
probe to each transmitted
particle using a translucent attack.
(2) Eve keeps the probes in a quantum memory  
(where non-orthogonal quantum states can be kept 
for long time~\cite{BHM}) 
till receiving {\em all} classical 
data including error-correction and privacy amplification data.
(3) Eve performs the optimal measurement on her probes in order to learn
the maximal information on the {\em final} key.
The case in which Eve attaches 
one probe (in a large-dimensional Hilbert-space)  
to all transmitted particles  
is called a {\em joint} or 
{\em coherent} attack~\cite{BBBSS},
and it is the most general possible attack.
No specific joint attacks were  
yet suggested; the collective attack defined above 
is the strongest 
joint attack suggested so far,
and there are good reasons
to believe that it 
is the strongest possible attack. 

The security of quantum cryptography is very complicated and tricky problem.
Several security claims done in the past were found later on to contain 
loopholes.
Recently, we become aware of three new such claims~\cite{DM2,oxford,LC}.
We hope that these approaches,
together with our approach 
really produce the solution;
yet it is important to have them all, since each of them 
has different advantages.

Our approach deals with 
error-correction and privacy amplification, by calculating the density matrices
which are available to the eavesdropper by the time all data transmissions
(classical and
quantum) are completed.
We provide an example of collective attacks based on
the ``translucent attack without entanglement'' of \cite{EHPP}, 
which leave Eve with probes in a pure state, and we prove security 
against them.
These attacks use the unitary transformation
$ {\cos \theta \choose \pm \sin \theta} \longrightarrow
                 {\cos \theta' \choose \pm \sin \theta'} 
                 {\cos \alpha \choose \pm \sin \alpha} $
with `$+$' for $\psi_0$, and `$-$'  
for $\psi_1$, where $\theta'$ is the angle of the states received
by Bob, and $\alpha$ is the angle of the states in Eve's hand.
The error-rate,
$p_e = \sin^2 (\theta - \theta')$,  
is the probability 
that Alice sent $\psi_0$ and Bob measured $\psi_0'$.
The connection between this induced error-rate and the angle $\alpha$
is calculated using the unitarity condition~\cite{EHPP}  
$\cos 2 \theta = \cos 2 \theta' \cos 2 \alpha$. For 
weak attacks which causes small error-rate the angle of Eve's probe
satisfies 
$\alpha = (p_e \tan^2 2\theta )^{1/4}$,
which is $(p_e)^{1/4}$ 
for $\theta = 22.5 \ {\rm deg}$.
In our case, the same translucent attack is 
performed on all the bits, 
and it
leaves Eve with $n$ probes, each in one of the two states ${c \choose \pm s}$, 
with $c = \cos \alpha$ and $s = \sin \alpha$.
As result, Eve holds an $n$ bits string $x$ which is concatenated from its bits
$(x)_1 \ (x)_2 \ldots (x)_n$.
For simplicity, we choose the final key to consist of one bit, 
which is the parity of the
$n$ bits.
Eve wants to distinguish between two density matrices corresponding
to the two possible values of this parity bit.
Our aim is to calculate the optimal mutual information she can 
extract from them. 

For our analysis we need some more notations. 
Let $\hat{n}(x)$ be the number of $1$'s in $x$. 
For two strings of equal length $x \odot y$ is the bitwise ``AND'', so that 
the bit $(x \odot y)_i$ is one if both $(x)_i$ and $(y)_i$ are one.
Also $x \oplus y$ is the bitwise  ``XOR'', so that 
$(x \oplus y)_i$ is zero if $(x)_i$ and $(y)_i$ are the same.
For $k$ (independent) strings,
$v_1 \ldots v_k$, of equal length let the set $\{v\}_k$
contain the $2^k$ linear combinations 
$(v_1), \ldots, (v_k), (v_1 \oplus v_1), (v_1 \oplus v_2), \ldots,
(v_1 \oplus v_2 \ldots \oplus v_k)$.
If these strings are not all different, 
then the original $k$ strings are linearly dependent.   
The quantum state of a string is the tensor product  
\begin{equation} \psi_x = {c \choose \pm s}{c \choose \pm s} \ldots
{c \choose \pm s} 
 = \left( \begin{array}{c}
      c c c  \ldots   c c c \\
 \pm  c c c  \ldots   c c s \\
             \ldots               \\
 \pm  s s s  \ldots   s s s \end{array} \right) \ , \label{value} \end{equation}
leaving in a $2^n$ dimensional Hilbert space. 
The sign of the $i$'th bit (in the middle expression) 
is plus for $(x)_i = 0$ and minus for
$(x)_i = 1$. 
The sign of the $j$'th term ($j = 0\ldots 2^{n-1}$)  
in the expression at the right depends on 
the parity of the string
$x \odot j$ and is equal to $(-1)^{p(x \odot j)}$.
The density matrix $\rho_x =  \psi_x   \psi_x^T$
also has for any $x$, the same terms up to the signs. 
We denote the absolute values by 
$\rho_{jk} \equiv |(\rho_x)_{jk}|$. 
The sign of each term $(\rho_x)_{jk}$ is given by
\begin{equation} 
                   (-1)^{p(x \odot j)}
                   (-1)^{p(x \odot k)} =
                   (-1)^{p[x \odot (j\oplus k)]}
\ . \label{sign} \end{equation}

A priori, all strings are equally probable 
and Eve needs to distinguish between the two density matrices
describing the parities.
%two density matrices:
%\begin{equation}
%\rho_0^{(n)}=\frac{1}{2^{n-1}}\sum_{x\,|\,p(x)=0}\!\!\!\!\!\rho_x\ ;
% \quad
%\ \rho_1^{(n)}=\frac{1}{2^{n-1}}\sum_{x\,|\,p(x)=1}\!\!\!\!\!\rho_x \ .
%\end{equation}
These matrices were calculated and analyzed in \cite{BMS} 
(henceforth, the BMS work), and independently in~\cite{DM} for the case
$\alpha = \pi/4$.
In case Eve is being told what 
the error-correction code is,
all strings consistent with the given error-correction 
code (the $r$ sub-parities) 
are equally probable,
and Eve need to distinguish between
the two density matrices: 
\begin{equation}
\rho_0^{(n,r)}=\frac{1}{2^{n-r-1}}\!\sum_{x\,|\,{{p(x)=0}\choose {x\ \!{\rm OECC}}}}\!\!\!\!\!\!\rho_x\ \!
; \quad \!\!\!
 \rho_1^{(n,r)}=\frac{1}{2^{n-r-1}}\!\sum_{x\,|\,{{p(x)=1}\choose {x\ \! {\rm OECC}}}}\!\!\!\!\!\!\rho_x 
\label{density-matrices}  \end{equation}
where ``OECC'' is a shortcut for obeys error-correction code.
Let us look at two simple examples where $n=5$,
one with $r=1$ and the second with $r=2$. 
Suppose that 
the parity of the first 
two bits, $(x)_1$ and $(x)_2$, 
is $p_1 = 0$.
Formally, this substring is described by the $n$-bit string 
$v_1 = 24$ which is $11000$ binary; The number of $1$'s in the first
two bits of a string $x$ is given by $\hat{n}(x \odot v_1)$,
and $x$ obeys the error-correction code if $p(x \odot v_1) = p_1$.
Let $v_d$ be the binary string ($11111$ in this case) 
which describes the substring
of the desired parity.
Eve could perform the optimal attack on
the three bits which are left, or in general, on $v_1 \oplus v_d$. 
For any such case, the optimal attack is given by the BMS work and the 
optimal information depends only on   
$\hat{n}(v_1 \oplus v_d)$, the {\em Hamming distance} between the two words.
This information (using eq. 53 of the BMS work) is  
\begin{equation} I(\hat{n}) = c {2k \choose k} \alpha^{2k} \label{BMS_info} \end{equation}
with $c=1$ for even $\hat{n}$ (which equals to $2k$) and $c= 1/ln2$ 
for odd $\hat{n}$ 
(that is $\hat{n} = 2 k - 1$). 
Suppose that Eve gets another parity bit $p_2 = 1$ of the binary string
$01100$ ($v_2 = 12$). 
Now, a string $x$ obeys the error-correction code if it also obeys
$p(x \odot v_2) = p_2$.  Clearly, it also satisfies 
$p[x \odot (v_1 \oplus v_2)] = p_1 \oplus p_2$.
In the general case there are $r$ independent parity strings,
and $2^r$ parity strings in the set $\{v\}_r$.  
The BMS result cannot be directly used but still provides some intuition: 
For each word (i.e., each parity string) 
$v_l \in \{v\}_r$, let  
$I(\hat{n} (v_l \oplus v_d))$ be the optimal information Eve could
obtain using eq.~\ref{BMS_info}. Also let 
$I_{sum}$ be the sum of these contributions from all such words.
In reality Eve cannot obtain $I_{sum}$
since each measurement changes
the state of the measured bits,
hence we expect that 
$I_{sum}$ bounds her
optimal information $I_{total}$ from above:
$I_{total} < I_{sum}$.
On the other hand, Eve knows all these words at once, 
and could take advantage of it,
thus we leave this as an unproven conjecture.  

In the following we find an explicit way to calculate
{\em exactly} the optimal
information. However, this exact result requires cumbersome calculations,
thus it is used only to verify the conjecture for short strings.

The parity of the full string is also known 
since the density matrix 
$\rho^{(n,r+1)}$ corresponds to either $\rho_0^{(n,r)}$ or $\rho_1^{(n,r)}$
depending on the desired parity $p_{r+1}$,
thus we add the string $v_{r+1}=v_d$.
There are $r+1$ 
independent sub-parities altogether, hence $2^{r+1}$ parity strings
in the set $\{v\}_{r+1}$. 
A string $x$ is included in $\rho^{(n,r+1)}$ if 
$p[x\odot v_l]=p_l$ for all given substring in $\{v\}_{r+1}$.
In the BMS work (where $r=0$) 
the parity density matrices were put in a block diagonal
form of $2^{n-1}$ blocks of size $2 \times 2$.
This result can be generalized to the case where $r$ parities of substrings
are given.
There will be $2^{n-r-1}$ blocks of size $2^{r+1} \times 2^{r+1}$.
We shall show that the $(jk)$'th term in a density matrix $\rho^{(n,r+1)}$
of $r+1$ sub-parities
is either zero,  $\rho_{jk}$ or $-\rho_{jk}$,
that is, either all the relevant strings contribute exactly the same term, or
half of them cancels the other half.
The proof can be skipped in a first reading. 
\begin{description}
\item[Theorem] 
\quad 

The element $(\rho^{(n,r+1)})_{jk}$ is zero if 
%\hfill\newline
$j\oplus k \not\in \{ v \}_{r+1}$,
and it is $\pm \rho_{jk}$ if 
$j \oplus k \in \{ v \}_{r+1}$.
\item[Proof] 
\quad 

In case $j\oplus k \not\in  
\{ v \}_{r+1}$
choose $C$ such that \hfill\newline
$ p[C\odot v_l] = 0 $ with all $(v_l)$'s in 
$\{ v \}_{r+1}$
and  \hfill\newline
$ p[C\odot (j \oplus k)] = 1 $
(many such $C$'s exists since $C$ has $n$ independent bits
and it need to fulfill only $r+2$ constraints).
For such a $C$ and for any $x$ which obeys the error-correction code
there exist one (and only one) $y$, $y = x \oplus C$,
which also obeys the code (due to the first demand) but has the opposite
sign in the $jk$'th element (due to the second demand),
so $(\rho_y)_{jk} = - (\rho_x)_{jk}$.
Since this is true for any relevant $x$, we obtain $(\rho^{(n,r+1)})_{jk} = 0$.

In case $j\oplus k \in \{ v \}_{r+1}$ such $C$ cannot exists,
and all terms must have the same sign: Suppose that there are two terms,
$x$ and $y$ with opposite signs. Then $C=x\oplus y$ satisfies the two
demands, leading to a contradiction.
\end{description}
This theorem tells us the place of all 
non-vanishing terms in the original ordering. 
The matrices can be reordered to a block-diagonal form
by exchanges of the basis vectors.
We group the vectors $s$, $s\oplus v_1$, etc., 
for all $(v_l)$'s in $\{v\}_{r+1}$
to be one after the other, so each such group 
is separated from the other groups. 
Now the theorem implies that all 
non-vanishing terms are grouped in blocks, and 
all vanishing terms are outside these blocks.
As result the matrix is
block-diagonal.
This forms $2^{n-r-1}$ blocks of size $2^{r+1} \times 2^{r+1}$.
All terms inside the blocks and their signs are given by eq.~\ref{value}
and~\ref{sign}  
respectively up to reordering.
The organization of the blocks depends only on the parity strings $v_l$ and
not on the parities $p_l$, thus,
$\rho_0^{(n,r)}$ and  
$\rho_1^{(n,r)}$ are block diagonalized in the same basis. 
The rank of a density matrix is the number of (independent) pure states which 
form it, and it is $2^{n-r-1}$ in case of the parity 
matrices (eq.~\ref{density-matrices}).
When these matrices are put in a block diagonal form, there are $2^{n-r-1}$
(all non-zero) blocks. Thus, the rank of each block is one, the corresponding
state is pure, and, when diagonalized, the non-vanishing term $a_j$ in 
the $j$'th block is the probability that a measurement will 
result in this block. 

In the BMS work ($r=0$), 
the information, in case of small angle, 
was found to be exponentially small with the length of the string.
When each probe is in a pure state, 
this result can be generalized to $r>0$ as follows:  
The optimal mutual information carried by two pure states
(in any dimension) is well known.
The two possible pure states in the $j$'th block of $\rho_0^{(n,r)}$
and $\rho_1^{(n,r)}$ can be written as 
$\cos\beta  \choose \pm \sin \beta $.
The optimal mutual information which can be obtained from the $j$'th block
is given by the overlap (the angle $\beta_j$) 
$ I_j = 1 + p_j \log p_j + (1-p_j) log (1-p_j)$, 
where $p_j = \frac{1 - \sin 2 \beta_j }{2}$;
The overlap 
is calculated using 
eq.~\ref{value} and~\ref{sign}.
Thus, for any given error-correction code, we can find 
the two pure states in each block, 
the optimal information $I_j$,
and finally, the total information 
$ I_{\rm total} = \sum_j a_j I_j $.
We did not use the value of $v_d$ in the proof, and thus, the 
final key could be the parity of any substring. 
Moreover, a similar method can be used to analyze keys
of several bits which can be formed from parities of several substrings.
 
We wrote a computer program which receives any (short)  
error-correction code 
and calculates the total information as a function of the angle
$\alpha$ between the pure states of the individual probes.
We checked many short codes (up to $n=8$) to
verify whether $I_{total} < I_{sum}$ as we conjectured.
Indeed, all our checks showed that the conjecture holds.
The information for small angle $\alpha$ is bounded 
by 
$ I_{sum} = C \alpha^{2k}$ as previously explained, where $C$ is given by 
summing the terms which contribute to the highest order of eq.~\ref{BMS_info},
and the Hamming distance $\hat{n}$ (which is $2k$ or $2k -1$), 
can be increased by choosing longer codes
to provide any desired level of security.

In addition to a desirable security level, the error-correction code 
must provide
also a desirable reliability;
A complete analysis must include also estimation of the probability
$p_f$ that Alice and Bob still has wrong (i.e. different) final key.
For enabling such analysis, one must use known error-correction codes.
Random Linear Codes allow for such analysis but cannot be used efficiently
by Alice and Bob.
Hamming codes~\cite{MS}, $H_r$ 
which use $r$ given parities for correcting one error in strings of length
$n=2^r - 1$, have  
an efficient decoding/encoding procedure
and a simple way to calculate $p_f$.  
An Hamming code has $2^r$ words in $\{v\}_r$, all of them,
except $00 \ldots 0$, are at the same distance
$\hat{n}=2^{r-1}-1$ from $v_d$.
Using our conjecture and eq.~\ref{BMS_info} 
(with $k=\frac{\hat{n}+1}{2}=2^{r-2}$) we obtain
$ I_{\rm total} < (2^r - 1) \frac{1}{\ln 2} {2^{r-1} \choose 2^{r-2} }
\alpha^{(2^{r-1})} + O\left(\alpha^{(2^r - 1)}\right) $.
For $r=3$ ($n=7$) this yields $I_{\rm total} < 60.6 \alpha^4$.
The exact calculation done using our computer program also 
gives the same
result, showing that the conjecture provides an
extremely tight bound in this case.
Using ${ 2^{r-1} \choose 2^{r-2} } < \frac{2^{(2^{r-1})}}{\sqrt(\frac{\pi}{2}
2^{r-1})}$ and some calculation
we finally obtain
\begin{equation} I_{\rm total} < \left( \frac{2}{\ln2 \sqrt\frac{\pi}{2}}\right)
\sqrt{2^{r-1}} (2\alpha)^{(2^{r-1})} \ , \end{equation}
bounding $I_{\rm total}$ to be exponentially small with $n$ 
[which follows from $2^{r-1} = (n+1)/2$].

The rate of errors in the string shared by Alice and Bob
(after throwing inconclusive results) is the normalized error-rate, 
$p_e^{{}^{(N)}} = p_e / (p_c + p_e)$,
where $p_c = \sin (\theta + \theta')$ is the probability of obtaining
a correct and conclusive result.
For small $\alpha$ it is 
$p_e^{{}^{(N)}} = \frac{2 p_e}{\sin^2 2 \theta} = 
\frac{2 \cos^2 2 \theta}{\sin^4 2 \theta} \alpha^4$.
The final error probability $p_f$ is given by
the probability to have more than one
error in the initial string, since the code corrects one error.
It is $p_f = \frac{n(n-1)}{2} (p_e^{{}^{(N)}})^2 + O[(n p_e^{{}^{(N)}})^3]$,
showing that we can use the Hamming codes as long as $n p_e^{{}^{(N)}} << 1$.
In case it is not, better codes such
as the BCH codes~\cite{MS} (which correct more than one error)
are required,  
but their analysis
is beyond the goals of this paper.

In conclusion, we presented new attacks on quantum key distribution schemes, 
directed against the final key, 
and we proved security against a specific one. 
This result, together with its extension to 
the analysis of probes in mixed state~\cite{future}, 
suggest that the optimal information
obtained by the {\em optimal} collective attack 
shall still show the same behavior as shown in our example.
Let us explain the intuition 
that the security against collective attacks 
implies security against any joint attack:
Most of the transmitted particles are not 
part of the $n$-bits string.
The correlations between the $n$ bits (as specified by the 
error-correction and privacy amplification)
as well as the random reordering of the bits  
are not known in advance. 
It is very reasonable that Eve can only lose 
by searching for such correlations when the particles are
transmitted through her. Thus, the best she can do is probe the particles
via the the best collective attack.

We are grateful to C. H. Bennett, G. Brassard, C. Cr\'epeau,
J. Smolin, A. Peres and the referees for many helpful discussion.
We are especially grateful to D. Mayers for his great help and many suggestions;
in particular for observing~\cite{DM1}
that $\rho_p^{(n,r)}$ are of a block diagonal form also for $r>0$  
(he proved it independently in another context~\cite{DM2}).
We also thank G. Brassard and the Universit\'e de 
Montr\'eal for hosting a productive meeting, which had an extremely
valuable 
contribution to this work.

%\end{multicols}
\end{document}